\documentclass[pra,aps,a4paper,noarxiv,onecolumn,floats,floatfix,nobibnotes,accepted=2022-10-20]{quantumarticle}

\usepackage[pdftex]{graphicx}
\usepackage[pdftex]{epsfig}
\usepackage{amsmath}
\usepackage{amssymb}
\usepackage{amsfonts}
\usepackage{color}
\usepackage{hhline}
\usepackage{tabularx}
\usepackage{upgreek}
\usepackage{soul}
\usepackage{float}
\usepackage{placeins}

\usepackage{array} % make column bigger (vertical)
\usepackage[colorlinks=true, allcolors=blue]{hyperref} %hyperrefs

\newcommand{\ee}{\mathrm{e}}
\newcommand{\I}{\mathrm{i}}

\newcommand{\OM}{\mathrm{OM}}
\newcommand{\OMC}{\mathrm{c}}
\newcommand{\laser}{\mathrm{l}}
\newcommand{\Mech}{\mathrm{m}}

\newcommand{\aux}{\mathrm{A}}

\newcommand{\forw}{\mathrm{fw}}
\newcommand{\backw}{\mathrm{bw}}

\newcommand{\STrip}{\mathrm{s}}

\newcommand{\kOM}{\kappa_\OMC}
\newcommand{\kOMi}{\kappa_\OMC^\mathrm{(i)}}
\newcommand{\kOMe}{\kappa_\OMC^\mathrm{(e)}}
\newcommand{\tkauxa}{{\tilde \kappa}_\aux^{(1)}}
\newcommand{\kauxa}{\kappa_\aux^{(1)}}
\newcommand{\kauxb}{\kappa_\aux^{(2)}}

\newcommand{\inp}{\mathrm{in}}
\newcommand{\inpe}{\mathrm{in,e}}
\newcommand{\inpi}{\mathrm{in,i}}

\newcommand{\outp}{\mathrm{out}}
\newcommand{\therm}{\mathrm{th}}

\newcommand{\dd}{\mathrm{d}}

\newcommand{\add}{\mathrm{add}}

\newcommand{\pulse}{\mathrm{p}}
\newcommand{\tmread}{\mathrm{s}}
\newcommand{\tmwrite}{\mathrm{g}}
\newcommand{\tmode}{\mathrm{tm}}

\begin{document}

\title{Coherent feedback in optomechanical systems in the sideband-unresolved regime}

\author{Jingkun Guo}
\author{Simon Gr\"oblacher}
\email{s.groeblacher@tudelft.nl}
\affiliation{Kavli Institute of Nanoscience, Department of Quantum Nanoscience, Delft University of Technology,\\ 2628CJ Delft, The Netherlands}

\begin{abstract}
    Preparing macroscopic mechanical resonators close to their motional quantum groundstate and generating entanglement with light offers great opportunities in studying fundamental physics and in developing a new generation of quantum applications. Here we propose an experimentally interesting scheme, which is particularly well suited for systems in the sideband-unresolved regime, based on coherent feedback with linear, passive optical components to achieve groundstate cooling and photon-phonon entanglement generation with optomechanical devices. We find that, by introducing an additional passive element -- either a narrow linewidth cavity or a mirror with a delay line -- an optomechanical system in the deeply sideband-unresolved regime will exhibit dynamics similar to one that is sideband-resolved. With this new approach, the experimental realization of groundstate cooling and optomechanical entanglement is well within reach of current integrated state-of-the-art high-Q mechanical resonators.
\end{abstract}

\maketitle

\section{Introduction}

Over the past decade, optomechanical systems have seen great progress towards studying fundamental physics and in realizing new applications~\cite{Stannigel2010,Krause2012,Marinkovic2018,Carlesso2019,Allain2020,Wallucks2020,Fiaschi2021,Westerveld2021}. In particular, microfabricated optomechanical systems with large mechanical resonators and an integrated optical cavity have attracted significant interest, as they provide a versatile and easy-to-use platform in many areas including sensing~\cite{Krause2012, Norte2018}, quantum networks~\cite{Bochmann2013, Cernotik2016, Arnold2020}, and for studying quantum effects in massive, macroscopic systems~\cite{Chen2013}. Demonstrating quantum effects in optomechanics has almost exclusively been the realm of systems in the sideband-resolved regime~\cite{Hofer2011,Paternostro2011,Palomaki2013,Aspelmeyer2014,Rakhubovsky2015,Marinkovic2018,Wallucks2020,Fiaschi2021}, with several notable exceptions~\cite{Rossi2018,Magrini2021,Chen2020}. This is in great part due to the strong suppression of either the phonon creation or the annihilation process through the optical cavity~\cite{Hofer2011,Aspelmeyer2014} when the cavity linewidth is small compared to the mechanical frequency. 

For fundamental tests using optomechanics, large and massive mechanical resonators are required however, which typically puts these systems far into the sideband-unresolved regime, where the mechanical resonance frequency is smaller than the linewidth of the optical cavity. While significant advancements have been made for device designs in the sideband-unresolved regime, both in terms of the mechanical resonator~\cite{Tsaturyan2017,Ghadimi2018,Guo2019,Beccari2021} and the integration with optical cavities with large optomechanical coupling strength~\cite{Leijssen2015,Guo2019,Guo2022}, the difference in suppression of the optomechanical sidebands is insignificant due to the large optical linewidth. This makes them incompatible with many of the standard approaches for quantum experiments used to date. At the same time, the large bandwidth of the optical cavity allows obtaining the information of the mechanical resonator and interacting with it very efficiently and with little delay. This has lead to several ideas specifically designed for this regime, including short-pulse~\cite{Vanner2011,Bennett2016,Khosla2017b,Clarke2020} and measurement-based~\cite{Genes2008} approaches for quantum state preparation. However, several challenges to experimentally implement these schemes exist, such as using an optical pulse with a duration much shorter than the mechanical oscillation period typically being limited by noise introduced by unwanted mechanical modes~\cite{Muhonen2019}. While measurement-based scheme are mostly focused on feedback cooling, creating photon-phonon entanglement through continuous measurement has also been proposed, with potential squeezing of the Einstein-Podolski-Rosen (EPR) quadratures of up to 50\%~\cite{Gut2020}.

In this work, by further exploring the large bandwidth of the optical cavity, and by using either a continuous laser or pulses that are much longer than the mechanical period, we propose schemes based on coherent feedback by external linear, passive optical elements. Effectively, the extra optical element creates an asymmetry in the suppression, similar to the sideband-resolved regime. We show that using this new approach, groundstate cooling and quantum entanglement, with squeezing of the EPR quadratures beyond 50\%, is possible with realistic experimental requirements, even with systems deep in the sideband-unresolved regime.

\section{Model}

\subsection{Optomechanical system}

We consider an optomechanical system consisting of a single mechanical mode and a single-mode optical cavity. The mechanical mode has an angular resonance frequency $\Omega_\Mech$ and an energy damping rate $\Gamma_\Mech$. Its field is bosonic, and the position and momentum quadratures are described by two normalized Hermitian operators $\hat X_\Mech$ and $\hat Y_\Mech$. The optical field has a resonance frequency of $\omega_\OMC$ and an energy damping rate $\kappa_\OMC$, with the amplitude and phase quadratures $\hat X_\OMC$ and $\hat Y_\OMC$. Throughout this work, we combine the quadratures $\hat u = (\hat X, \hat Y)^T$ to simplify our expressions. The quadratures satisfy the commutation relation $[\hat u_{\alpha, i}, \hat u_{\beta, j}] = \I \delta_{\alpha \beta} \varepsilon_{ij}$, where $\alpha, \beta$ are for $\OMC$ (cavity field) or $\Mech$ (mechanical field), $i, j \in \{1, 2\}$ for the $X$ or $Y$ quadrature. $\varepsilon_{ij}=1$ for $i=1,~j=2$, $\varepsilon_{ij}=-1$ for $i=2,~j=1$, and 0 otherwise. The annihilation operators of the two bosonic fields are $\hat a_\alpha = (\hat X_\alpha + \I Y_\alpha) / \sqrt{2}$. We use the the frame rotating with the laser (drive) frequency $\omega_\laser$, and we define the detuning $\Delta_\OMC = \omega_\laser - \omega_\OMC$ as being the frequency difference between the input laser and the cavity field. The mechanical resonator and the optical cavity couple dispersively, with a (linearized) coupling strength $g=\sqrt{n_\OMC} g_0$~\cite{Aspelmeyer2014,Bowen2015}, where $g_0$  is the single photon coupling rate enhanced by the intra-cavity photon number $n_\OMC$.

The mechanical resonator and the optical cavity couple to the environment through their respective loss channels, the energy dissipation rate $\Gamma_\Mech$ and $\kOM$. For the mechanical mode, in a typical experiment in the sideband-unresolved regime at temperature $T$, the thermal phonon excitation is given by $n_\therm \approx \frac{k_\mathrm{B} T}{\hbar \Omega_\Mech} \gg 1$, with $k_\mathrm{B}$ being the Boltzmann constant. We further assume the mechanical quality factor $Q_\Mech$ to be large. The bath only couples to the momentum quadrature of the harmonic oscillator and is approximately Markovian, $\hat u_\Mech^\inp = (0,~ {\hat Y}_\Mech^\inp)$, with $\left\langle {{\hat Y}_\Mech^\inp (t) {\hat Y}_\Mech^\inp (t^\prime)} + {{\hat Y}_\Mech^\inp (t^\prime) {\hat Y}_\Mech^\inp (t)} \right\rangle \approx (n_\therm + 1/2) \delta (t - t^\prime)$~\cite{Gut2020}. Due to the high frequency of the cavity field any thermal excitations can be neglected and the cavity input field is in the vacuum state. We consider two loss channels for the cavity, where one is due to the coupling to an external mode with an energy dissipation rate of $\kOMe$ and all other losses are included in $\kOMi$, with $\kappa_\OMC=\kOMe+\kOMi$. The associated optical field are $\hat u_{\OMC}^{\inpe}$ and $\hat u_{\OMC}^{\inpi}$, respectively. The linearized dynamics of the system are then described by the Quantum Langevin equation~\cite{Bowen2015,Gut2020}
\begin{equation}\label{eq:OM_lin_Langevin}
    \begin{gathered}
        \dot {\hat X}_\OMC = - \frac{\kappa_\OMC}{2} \hat X_\OMC - \Delta_\OMC \hat Y_\OMC + \sqrt{\kOMe} \hat X_\OMC^{\inpe} + \sqrt{\kOMi} \hat X_\OMC^{\inpi}, \\
        \dot {\hat Y}_\OMC = \Delta_\OMC \hat X_\OMC - \frac{\kappa_\OMC}{2} \hat Y_\OMC - 2 g X_\Mech + \sqrt{\kOMe} \hat Y_\OMC^{\inpe} + \sqrt{\kOMi} \hat Y_\OMC^{\inpi}, \\
        \dot {\hat X}_\Mech = \Omega_\Mech \hat Y_\Mech, \\
        \dot {\hat Y}_\Mech = - \Omega_\Mech \hat X_\Mech - \Gamma_\Mech Y_\Mech - 2 g X_\OMC + \sqrt{2 \Gamma_\Mech} \hat Y_\Mech^{\inp}.
    \end{gathered}
\end{equation}
Only the field coupled back to the external mode can be collected, $\hat u_\OMC^{\outp} = \hat u_\OMC^{\inpe} - \sqrt{\kOMe} u_\OMC$.

\subsection{Coherent feedback with linear optical elements}

\begin{figure}[!t]
    \centering
    \includegraphics[width=1.\columnwidth]{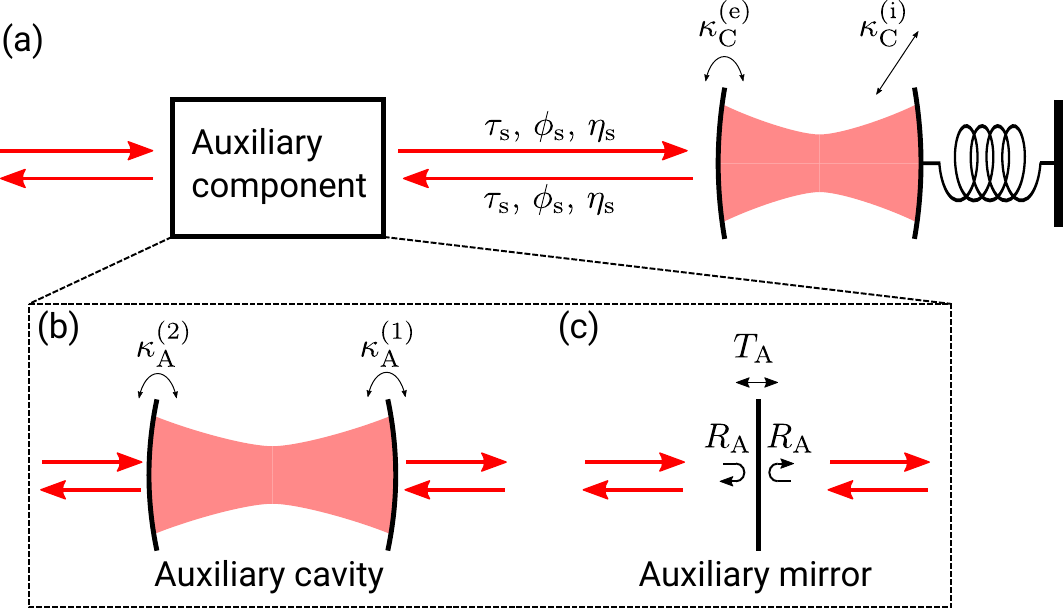}
    \caption{(a) Coherent feedback scheme. The optomechanical cavity, with an intrinsic loss rate $\kOMi$ and an external coupling $\kOMe$, is connected to an auxiliary component via an optical path. The output light from the optomechanical cavity couples to the auxiliary component, and then travels back to the optomechanical cavity, forming a feedback loop. The optical path introduces a single-way delay of $\tau_\STrip$, phase $\phi_\STrip$, with a single-way efficiency of $\eta_\STrip$ and can be used as a channel to couple driving laser into the feedback system and to perform measurements. We consider (b) an auxiliary cavity or (c) an auxiliary mirror for the auxiliary component in this work. The auxiliary cavity has a coupling rate $\kauxa$ to the internal feedback optical path and $\kauxb$ to the outside. The reflectivity of the auxiliary mirror is $R_\aux$.}
    \label{fig:GeneralScheme}
\end{figure}

With additional resources included, coherent feedback can have a significant impact on the dynamics of a system~\cite{Yanagisawa2006,James2008,Hamerly2012,Yamamoto2014,Combes2017, Ojanen2014,Bennett2014,Karg2019,Li2017,Feng2017,Wang2017b,Lau2018,Karg2020,Harwood2021,Schmid2022}. In our approach, we consider the optomechancial cavity being connected to either an external optical cavity or a mirror via an optical path, as shown in Figure~\ref{fig:GeneralScheme}. Experimentally, the optical path might be realized by free-space optics, an optical fiber, or an on-chip waveguide. The light traveling through the path acquires a constant single way delay $\tau_\STrip$ and a phase shift $\phi_\STrip$, and the path has a single way efficiency $\eta_\STrip$. The input and output of the optomechanical cavity are related to the input and output of the auxiliary component,
\begin{equation}
    \begin{gathered}
        \hat u_\aux^{\inp,1} (t) = \sqrt{\eta_\STrip} R(\phi_\STrip) \hat u_\OMC^{\outp} (t-\tau_\STrip) + \sqrt{1 - \eta_\STrip} \hat u_\forw^\inp (t), \\
        \hat u_\OMC^\inp (t) = \sqrt{\eta_\STrip} R(\phi_\STrip) \hat u_\aux^{\outp,1} (t-\tau_\STrip) + \sqrt{1 - \eta_\STrip} \hat u_\backw^\inp (t).
    \end{gathered}
\end{equation}
$\hat u_\forw^\inp$ and $\hat u_\backw^\inp$ are the input vacuum field due to the loss in the optical connection, and the subscript $\aux$ denotes the field of the auxiliary component. $R$ is a rotational matrix,
\begin{equation}\label{eq:RotMat}
    R(\phi) = 
    \begin{pmatrix}
        \cos \phi & \sin \phi \\  -\sin \phi & \cos \phi
    \end{pmatrix},
\end{equation}
due to the phase acquired in the optical path. The auxiliary component in general has two sides, with channel 1 coupling to the optomechanical system, and channel 2 coupling to the outside, which can be used for driving and readout. With a constant drive field from channel 2, the input $\hat u_\aux^{\inp,2}$ is a vacuum field since only the fluctuations are considered.

To simplify our discussion, we will mostly focus on using an optical cavity as the auxiliary component. The intracavity field is then denoted as $\hat u_\aux = (\hat X_\aux, ~ \hat Y_\aux)$ and
\begin{equation}
    \begin{gathered}
        \dot {\hat X}_\aux = - \frac{\kappa_\aux}{2} \hat X_\aux - \Delta_\aux \hat Y_\aux + \sqrt{\kauxa} \hat X_\aux^{\inp,1} + \sqrt{\kauxb} \hat X_\aux^{\inp,2}, \\
        \dot {\hat Y}_\aux = \Delta_\aux \hat X_\aux - \frac{\kappa_\aux}{2} \hat Y_\aux + \sqrt{\kauxa} \hat Y_\aux^{\inp,1} + \sqrt{\kauxb} \hat Y_\aux^{\inp,2},
    \end{gathered}
\end{equation}
where $\kauxa$ and $\kauxb$ are the loss rates, or coupling, to channel 1 and 2 and the total loss rate of the cavity $\kappa_\aux = \kauxa+\kauxb$. The output on both channels is given by the input-output relation~\cite{Aspelmeyer2014}
\begin{equation}
    \begin{gathered}
        \hat u_\aux^{\outp,k} = \hat u_\aux^{\inp,k} - \sqrt{\kappa_\aux^{(k)}} \hat u_\aux,
    \end{gathered}
\end{equation}
where $k \in \{1, 2\}$ denotes the index of the coupling channel.

For the coherent feedback cooling, we will also explicitly consider the scheme where a mirror with a reflectivity $R_\aux$ is used as the coherent feedback component. In this case, there are no addition fields with a time derivative, and the output is directly given by the input
\begin{equation}
    \hat u_\aux^{\outp,1} = \sqrt{R_\aux} u_\aux^{\inp,1} + \sqrt{1 - R_\aux} u_\aux^{\inp,2}.
\end{equation}
Here, we drop an added phase from the reflection as any additional phases can be included into the phase of the optical path $\phi_\STrip$. Combining the dynamics in the system, we obtain a Langevin equation of the form,
\begin{equation}\label{eq:CFBLangevinForm}
    \begin{aligned}
        & \dot {\hat u} (t) + D \dot {\hat u} (t - \tau) \\
        &\quad = A_0 \hat u(t) + A_1 u (t - \tau) + \sum_{n=0}^{2} C_n \hat u_\inp(t - n \tau_\STrip),
    \end{aligned}
\end{equation}
$D$ and $A_n$ ($n=0,1$) define the interaction between fields in different elements in the system, while the $C_n$ matrices give the coupling to the external fields. The delayed response are given in the matrices $D$ and $A_n$, $C_n$ with $n \neq 0$. All the localized fields are included in $\hat u$, and all the input fields are included in $\hat u_\inp$. For example, for the coherent feedback with an auxiliary cavity, we can write $\hat u = (\hat X_\Mech,~ \hat Y_\Mech,~ \hat X_\OMC,~ \hat Y_\OMC,~ \hat X_\aux,~ \hat Y_\aux)^T$. The delay $\tau$ depends on the scheme, which for the coherent feedback with a mirror is $\tau = \tau_\STrip$ and with an auxiliary cavity $\tau = 2 \tau_\STrip$.

Equation~\eqref{eq:CFBLangevinForm} is a delay differential equation, which is hard to solve in general. However, for coherent feedback cooling the system will reach a steady state and with the stability test~\cite{Louisell2001,Olgac2004}, this steady state can be solved in the Fourier domain and the final phonon occupancy can be obtained (Appendix~\ref{sect:SteadyState}). For entanglement generation and verification, we will consider the special case with $\tau_\STrip = 0$, for which the solution can be obtained by solving the time evolution of the covariance matrix.

\section{Results}

\subsection{Similarity of a sideband-resolved system with auxiliary cavity}
\label{ss:SimSBR}

Let us first consider a simplified but illustrative model, which can be solved analytically. The optomechanical system is in the deep sideband-unresolved regime, $\kappa_\OMC \gg \Omega_\Mech$. The cavity field has a dynamics that is much faster than the dynamics of the mechanical resonator. It is therefore possible to eliminate the derivative to the cavity field by approximating it with an instant response to the mechanical resonator~\cite{Genes2008,Gut2020}. If we choose $\Delta_\OMC=0$ the optical field in equation~\eqref{eq:OM_lin_Langevin} is then given by
\begin{equation}
    \begin{gathered}
        \frac{\kappa_\OMC}{2} \hat X_\OMC \approx \sqrt{\kOMe} \hat X_\OMC^{\inpe} + \sqrt{\kOMi} \hat X_\OMC^{\inpi}, \\
        \frac{\kappa_\OMC}{2} \hat Y_\OMC \approx - 2 g X_\Mech + \sqrt{\kOMe} \hat Y_\OMC^{\inpe} + \sqrt{\kOMi} \hat Y_\OMC^{\inpi}. \\
    \end{gathered}
\end{equation}
For the feedback part, we consider $\phi_\STrip$ = 0. This can be achieved experimentally by locking the length of the feedback path to a fixed value. We further choose a short feedback path $\tau_\STrip \ll 2\pi / \Omega_\Mech$, allowing us to approximate $\tau_\STrip \approx 0$. This results in a linearized Langevin equation
\begin{widetext}
\begin{equation}\label{Eq:sim_single_cav_Langevin}
    \begin{gathered}
        \dot {\hat X}_\aux = \frac{\tilde \kappa_\aux}{2} \hat X_\aux - \Delta_\aux \hat Y_\aux + \sqrt{\tkauxa} \hat {\tilde X}_\aux^{\inp,1} + \sqrt{\kauxb} \hat X_\aux^{\inp,2}, \\
        \dot {\hat Y}_\aux = \Delta_\aux \hat X_\aux + \frac{\tilde \kappa_\aux}{2} \hat Y_\aux - 2 \tilde g \hat Y_\Mech + \sqrt{\tkauxa} \hat {\tilde Y}_\aux^{\inp,1} + \sqrt{\kauxb} \hat Y_\aux^{\inp,2}, \\
        \dot {\hat X}_\Mech = \Omega_\Mech \hat Y_\Mech, \\
        \dot {\hat Y}_\Mech = - \Omega_\Mech \hat X_\Mech - \Gamma_\Mech \hat Y_\Mech + \sqrt{2\Gamma_\Mech} P_\inp - 2 \tilde g \left( \hat X_\aux - \sqrt{\frac{(1-\eta_\STrip) \xi_1}{\eta_\OM \eta_\STrip \kauxa}} \hat X_\add \right),
    \end{gathered}
\end{equation}
\end{widetext}
where $\eta_\OM = \kOMe/\kappa_\OMC$, and $\xi_n = 1 - \eta_\STrip^n r_\OM^n$ with $r_\OM = 1 - \frac{\kOMe}{\kappa_\OM/2}$. 
We have introduced two effective bosonic fields, $\hat {\tilde u}_\aux ^{\inp,1}$ and $\hat u_\add$, whose full expressions are given in Appendix~\ref{app:eff_field_sim_single_cav}. Equations~\eqref{Eq:sim_single_cav_Langevin} results in dynamics that are similar to a bare optomechanical cavity where the mechanical resonator is directly coupled to the auxiliary cavity when compared to equation~\eqref{eq:OM_lin_Langevin}, with modified parameters,
\begin{equation}
    \begin{gathered}
        \tkauxa = \frac{\xi_2}{\xi_1^2} \kauxa, \\
        \tilde \kappa_\aux = \tkauxa + \kauxb, \\
        \tilde g = - \frac{\sqrt{\eta_\STrip \kauxa \kOMe}}{\xi_1 \kappa_\OMC/2} g.
    \end{gathered}
\end{equation}
Now, the effective optical decay rate is $\tilde \kappa_\aux$. By using a narrow-linewidth auxiliary cavity, it therefore effectively brings the system into the sideband-resolved regime. In particular, with the optical cavity of the optomechanical system being overcoupled, which can be routinely achieved experimentally~\cite{Guo2019,Krause2015,Guo2022}, $r_\OMC<0$ and $\xi_2/\xi_1^2 < 1$, $\tilde \kappa_\aux < \kappa_\aux$. The effective optical decay rate is smaller than the actual decay rate of the auxiliary, enabling even less stringent linewidth requirements for the auxiliary cavity. Experimentally, external Fabry-P\'{e}rot cavities can have a much narrower linewidth than the mechanical frequencies in many integrated optomechanical systems. Connecting to an external cavity, however, introduces an additional delay due to the optical path length, which can be significant for mechanical resonators with high frequencies~\cite{Eichenfield2009a,Leijssen2015,Arnold2020}. In this regime, an on-chip auxiliary optical cavity~\cite{Wu2020b,Puckett2021} is able to provide both small linewidth and short optical path to ensure $\tau_\STrip \ll 2\pi\Omega_\Mech$.

The drastic reduction of the effective linewidth of the optical cavity, from $\kappa_\OMC$ to $\tilde \kappa_\aux$, allows to now realize experiments and applications originally proposed for the sideband-resolved regime, with systems with broad integrated cavities or very low mechanical frequencies. This however comes at the expense of a reduction in the (effective) optomechanical coupling rate $\tilde g$. Furthermore, the added noise $\hat X_\add$ increases the effective phonon number of the bath,
\begin{equation}
    \tilde n_\therm = n_\therm + 4 \frac{1-\eta_\STrip}{\xi_1} \frac{g^2}{\Gamma_\Mech \kappa_\OMC}.
\end{equation}
Note that the added noise vanishes at the limit of $\eta_\STrip \rightarrow 1$, i.e., no loss in the optical path, while it does not require a fully overcoupled optomechanical cavity ($\kOMi = 0$).

To better understand the effect of the coherent feedback, we consider another important figure of merit in optomechanics, the quantum cooperativity $C_\mathrm{qu} = \frac{4 g^2}{n_\therm \kappa \Gamma_\Mech}$~\cite{Aspelmeyer2014}. Larger $C_\mathrm{qu}$ means a more robust optomechanical interaction compared to the photon and phonon decoherence and is thus favorable for optomechanical experiments in the quantum regime~\cite{Aspelmeyer2014}. The quantum cooperativity with the coherent feedback is then given by
\begin{equation}
    \begin{aligned}
        \tilde C_\mathrm{qu} = 
        \frac{4 \tilde g^2}{\tilde \kappa_\aux \tilde \Gamma_\Mech \tilde n_\therm}
        =
        \frac{4 \eta_\STrip \eta_\OMC (\kauxa/\kappa_\aux)  /\xi_1^2}{\left( 1 + 2\frac{\eta_\STrip r_\OMC}{\xi_1} \frac{\kauxa}{\kappa_\aux} \right) \left( 1 + \frac{1-\eta_\STrip}{\xi_1} C_\mathrm{qu} \right)} C_\mathrm{qu},
    \end{aligned}
\end{equation}
For a lossless optical path, the effective quantum cooperativity is enhanced as long as $\kauxa > \kappa_\aux/2$, i.e., the auxiliary cavity is overcoupled to the optomechanical cavity. This is due to the fact that an overcoupled auxiliary cavity recycles photons. With an inefficient feedback $\eta_\STrip < 1$, the enhancement is reduced when the original quantum cooperativity is very large due to the added noise from the optical path.

\begin{figure}[!t]
 \centering
  \includegraphics[width=1.\columnwidth]{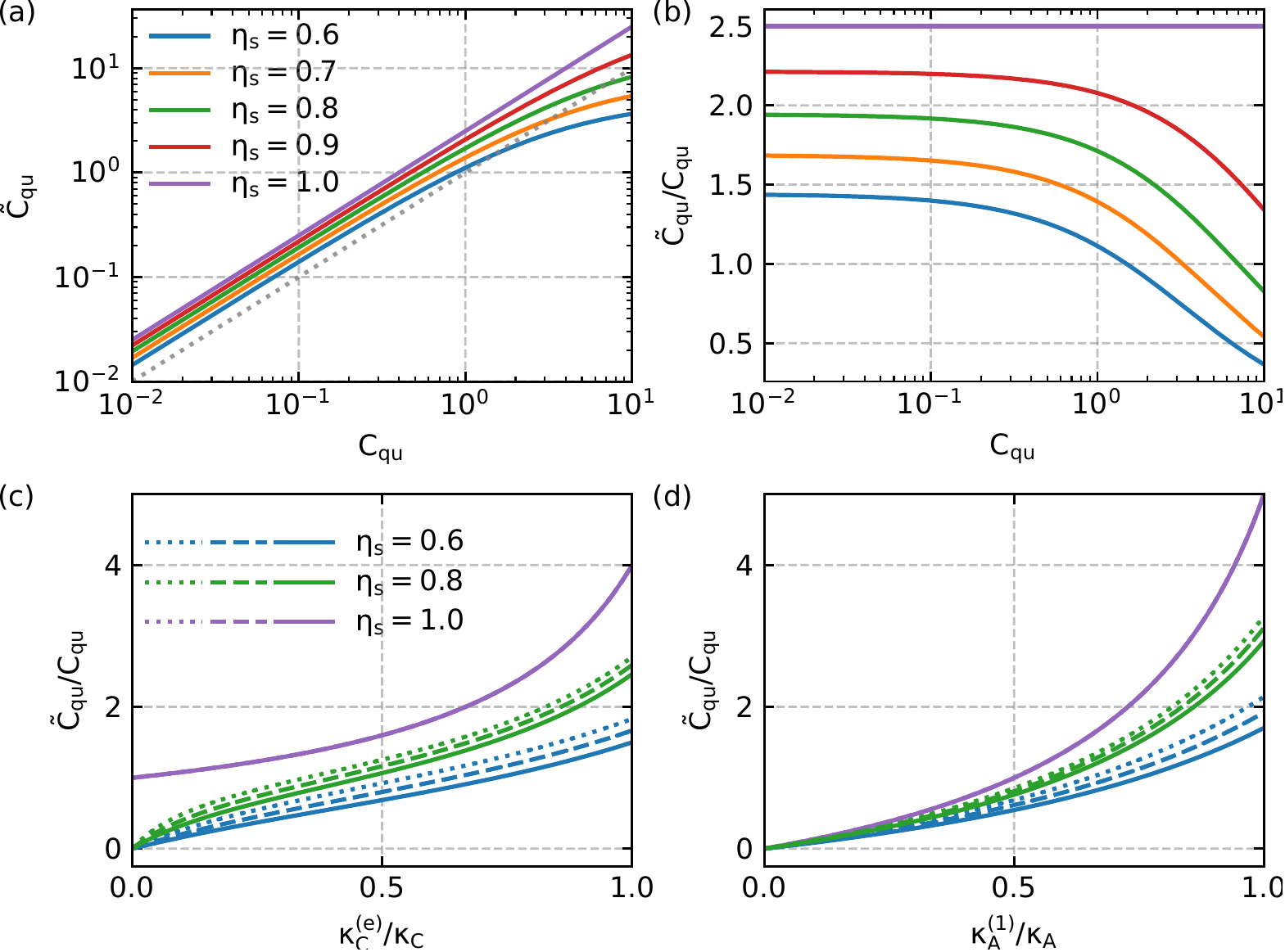}
  \caption{(a) Effective quantum cooperativity as a function of the original quantum cooperativity and (b) their ratio at different single-way optical path efficiency $\eta_\STrip$. When the original quantum cooperativity is large, the enhancement can be reduced. The dotted line in (a) is for $\tilde C_\mathrm{qu} = C_\mathrm{qu}$. Note that (a) and (b) share the same legend. (c,d) The quantum cooperativity ratio as a function of the coupling of the optical cavities to the feedback system. The dotted line, dashed lines, and the solid line are for an original quantum cooperativity of 0.1, 0.5 and 1, respectively, and the color labels are indicated in (c). In (c), $\kappa_\OMC/2\pi = 10$ GHz is fixed. In (d), $\kappa_\aux/2\pi = 500$ kHz is fixed. The other parameters for the plot are listed in the appendix.}
  \label{fig:EffCqu}
\end{figure}

For a set of practical parameters (see Appendix~\ref{sect:DefaultParams}), the effective quantum cooperativity and the ratio between the effective and the original quantum cooperativity are shown in Figure~\ref{fig:EffCqu}(a,b). Without loss in the optical path, the cooperativity is enhanced by up to a factor of 1.5. Despite the added noise, we note that the enhancement is robust against loss in the feedback path. With a moderate single way efficiency of 70\%, the enhancement of the quantum cooperativity persists until $C_\mathrm{qu} \approx 3.25$, which allows for experiments with a relatively large quantum cooperativity. This regime is crucial for many applications at high temperature, such as cooling and non-classical state generation~\cite{Chan2011,Gut2020,Ren2020}. Improving the efficiency drastically increases the enhancement region, with a single-way efficiency of 0.8 yielding an upper bound of $C_\mathrm{qu} = 6.98$, more than doubling the previous value. The results for changing the coupling efficiency $\kOMe/\kappa_\OMC$ and $\kauxa/\kappa_\aux$, while fixing $\kappa_\aux$ and $\kappa_\aux$, are plotted in Figure~\ref{fig:EffCqu}(c,d). As expected, increasing the coupling to the internal feedback system increases the quantum cooperativity enhancement.

\subsection{Coherent feedback cooling}\label{ss:Results:Cooling}

Preparing a mechanical resonator close to its quantum groundstate has been a major driving force in optomechanics over the past decades~\cite{OConnell2010,Chan2011,Teufel2011b,Whittle2021}, enabling many quantum experiments~\cite{Barzanjeh2022}. Inspired by the similarity between a system in the sideband-unresolved regime with coherent feedback and a sideband-resolved system, we show that it is possible to reduce the phonon occupation with our coherent feedback scheme. With large quantum cooperativity even groundstate cooling can be achieved. Furthermore, the additional parameters available to control the optomechanical system, allows it to perform better than a similar optomechanical system without the coherent feedback.

\begin{figure}[!t]
    \centering
    \includegraphics[width=1.\columnwidth]{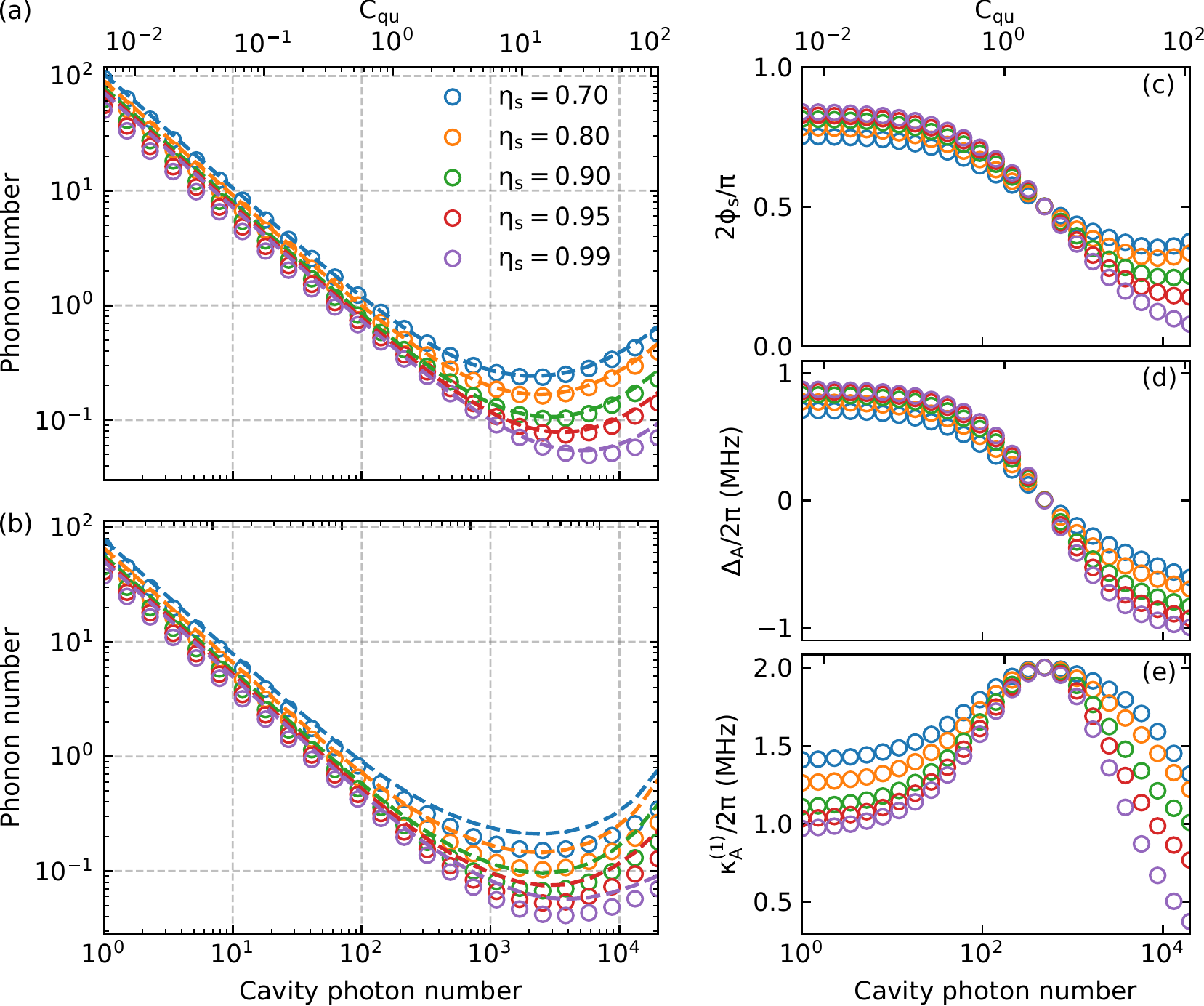}
    \caption{Coherent feedback cooling with an auxiliary cavity without delay. The dashed lines show the sideband cooling result for a similar optomechanical cavity (see section \ref{ss:SimSBR}), and circles show the optimized results. (a) Only $\phi_\STrip$ and $\Delta_\aux$ are optimization parameters and while in (b) $\phi_\STrip$, $\Delta_\aux$ and $\kappa_\aux^{(1)}$ are optimization parameters. The optimized parameters for (b) are plotted in (c-e). The optimized parameters for (a) are plotted in Figure~\ref{fig:Cooling_CR_tau_0_supp}.}
    \label{fig:Cooling_CR_tau_0}
\end{figure}

Figure~\ref{fig:Cooling_CR_tau_0}(a) shows the average phonon occupancy for an optomechanical system that is coupled to an auxiliary cavity without delay. The phonon number is minimized numerically with respect to the phase acquired on the optical path $\phi_\STrip$ and the detuning of the auxiliary cavity $\Delta_\aux$. When the quantum cooperativity approaches 1, the phonon number starts to drop below 1. For a single-way efficiency of 70\%, 80\% and 90\%, it is possible to achieve a minimum phonon number of 0.24, 0.16 and 0.10, respectively. When the quantum efficiency becomes too large, the added noise from the feedback and the back-action noise in the optomechanical system dominates and thus the phonon number increases. The sideband cooling results of an equivalent optomechanical system (Equation~\eqref{Eq:sim_single_cav_Langevin}) is shown as well. They correspond to a feedback cooling where $\phi_\STrip = 0$ and $\Delta_\aux=-\Omega_\Mech$. We see that the optimized phonon number is slightly lower and that the reduction is larger when the efficiency is higher. A greater benefit can be obtained when we further include $\kauxa$ as an optimization parameter (cf.\ Figure~\ref{fig:Cooling_CR_tau_0}b). As the auxiliary cavity is not part of the optomechanical system, we can freely choose its parameters as long as they are experimentally feasible. Here, we keep the coupling to the outer channel the same as before, $\kauxb/2\pi=100$~kHz and for a single-way efficiency of 70\%, 80\% and 90\%, it is then possible to achieve a minimum phonon number of 0.15, 0.10 and 0.07, respectively. The optimized parameters are plotted in Figure~\ref{fig:Cooling_CR_tau_0}(c-e), showing the optimal parameters being very different from simply setting $\phi_\STrip = 0$ and $\Delta_\aux=-\Omega_\Mech$. The optimal detuning is not necessarily on the red sideband due to the extra degree of freedom $\phi_\STrip$. The optimal linewidth is larger than or comparable to the mechanical frequency, resulting in an auxiliary cavity that is heavily overcoupled to the feedback system. It is also interesting to note that we find an optimal point that is fixed for different efficiencies over the feedback optical path. It occurs at roughly 500 intra-cavity photons ($C_\mathrm{qu}\approx 2.8$), with the optimal set of parameters $\phi_\STrip \approx \pi/4$, $\Delta_\aux/2\pi \approx 0$~MHz, and $\kauxa/2\pi \approx 2$~MHz. The reason for this fixed point is not yet understood and it may require solving the model analytically.

\begin{figure}[!t]
    \centering
    \includegraphics[width=1.\columnwidth]{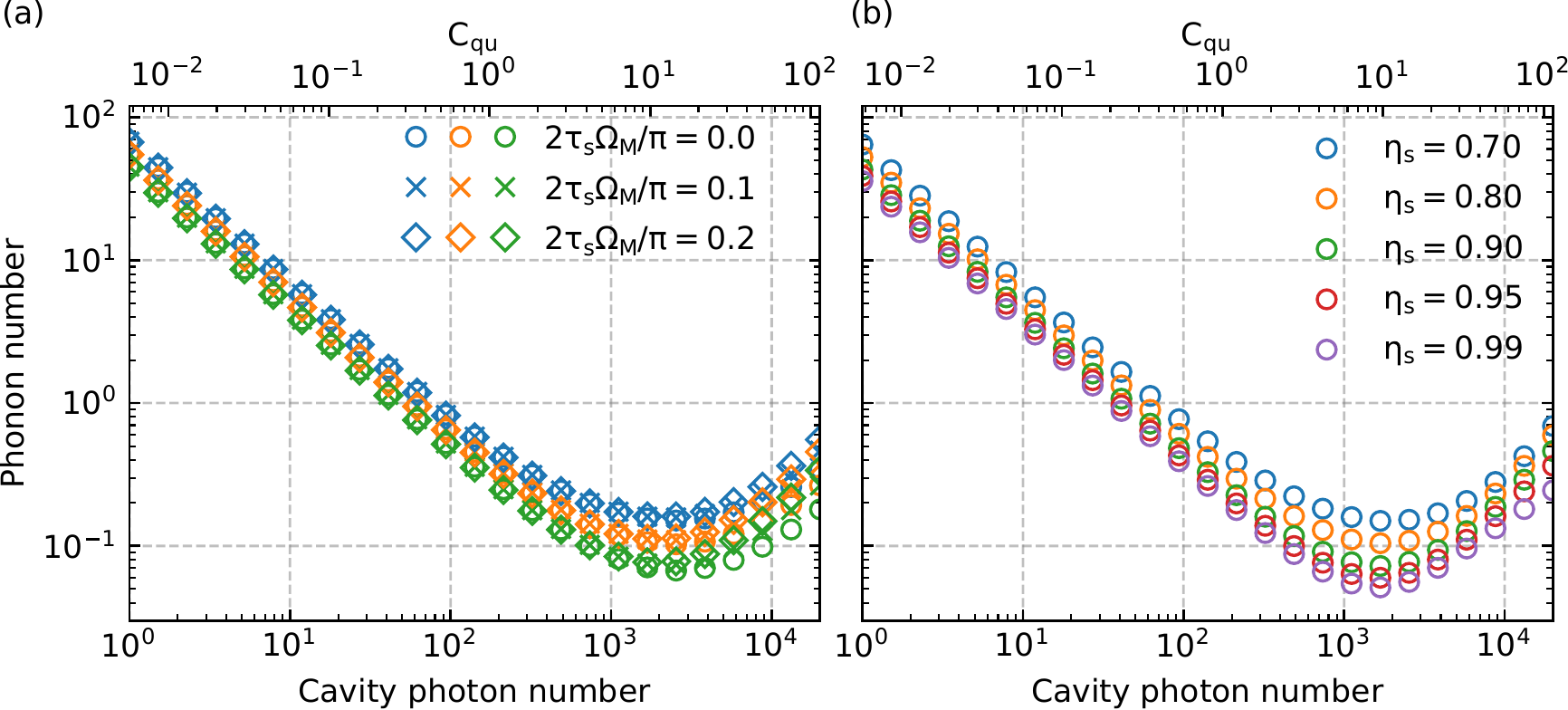}
    \caption{Coherent feedback with delay for (a) an auxiliary cavity and (b) an auxiliary mirror. In (a), the round trip delay is 0.0, 0.05 and 0.1~$\upmu$s, corresponding to a feedback phase $2 \Omega_\Mech \tau_\STrip$ of 0, $0.1\pi$ and $0.2 \pi$. The single way efficiency over the optical path is $\eta_\STrip = 0.7,~0.8,~0.9$ for blue, orange and green curves. The phonon number is minimized numerically with respect to $\kauxa$, $\Delta_\aux$ and $\phi_\STrip$. In (b), the reflectivity of the auxiliary mirror is set to 1. Non-unity reflectivity can be included in the optical path efficiency. The phonon number is minimized with respect to $\tau_\STrip$ and $\phi_\STrip$.}
    \label{fig:Cooling_CR_MR_sweep_tau}
\end{figure}

In an actual experiment, the optical path will always introduce a delay, which reduces the coherence of the feedback. For low frequency mechanical resonators, this delay can however be kept small compared to the oscillation period. We would like to stress that the delay does not introduce a significant reduction of the cooling performance if the mechanical decoherence rate is small, i.e., the decoherence has a time scale that is much longer than the feedback delay. In Figure~\ref{fig:Cooling_CR_MR_sweep_tau}(a), the cavity-assisted coherent feedback with delay at different $\eta_\STrip$ is studied, where the round trip delay $2\tau_\STrip$ is set to 0, 0.05 and 0.1~$\upmu$s. This corresponds to a single-way free-space optical path of 0, 7.5, and 15~m, respectively. No significant higher phonon number for different delays with different $\eta_\STrip$ can be observed. The cooling performance is worse at very large photon number, which is however already outside the optimal regime.

\begin{figure}[!t]
    \centering
    \includegraphics[width=0.8\columnwidth]{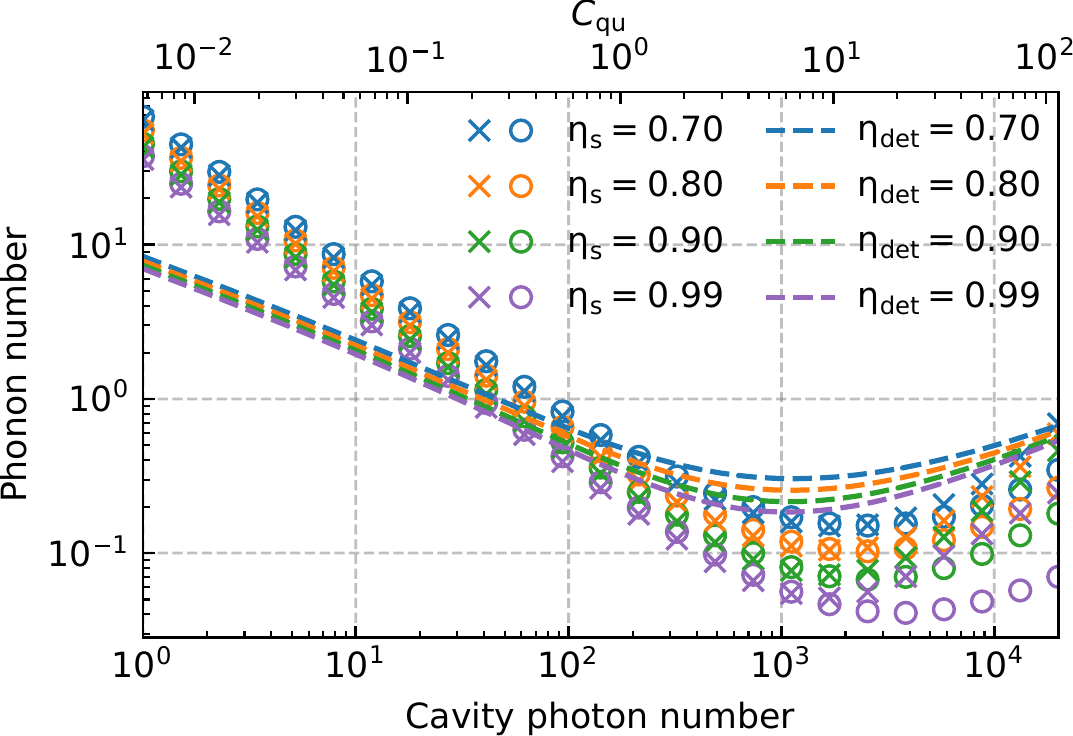}
    \caption{Comparison between coherent feedback by mirror (crosses, results from \ref{fig:Cooling_CR_tau_0}(b)), by auxiliary cavities (circles, results from \ref{fig:Cooling_CR_MR_sweep_tau}(b)), and from measurement-based feedback (dashed lines, results calculated from~\cite{Genes2008}). The measurement based feedback cooling has a detection efficiency $\eta_\mathrm{det}$ that is equal to $\eta_\STrip$ of the corresponding coherent feedback case.}
    \label{fig:Cooling_CFB_Compare_MFB}
\end{figure}

Besides the coherent feedback with an auxiliary cavity, coherent feedback cooling can also be achieved by using a mirror as the feedback element (cf.\ Figure~\ref{fig:GeneralScheme}(c)) in the presence of optical delay. With a round-trip delay of around $1/4$ of the mechanical oscillation cycle, the feedback can introduce a significant damping force onto the mechanical resonator when $\phi_\STrip$ is properly tuned, which we show in Figure~\ref{fig:Cooling_CR_MR_sweep_tau}(b). For simplicity, we assume a mirror reflectivity of 1, while any deviation from unity can be directly included in the optical path efficiency $\eta_\STrip$. We find a similar cooling performance compared to the coherent feedback with an auxiliary cavity. At large quantum cooperativity, the resulting phonon number is slightly higher due to the larger delay which introduces an additional incoherent signal to the feedback. However, similar to the feedback with the cavity, the impact is small at the optimal quantum cooperativity. A direct comparison is plotted in Figure~\ref{fig:Cooling_CFB_Compare_MFB}.

Additonally, we compare the coherent feedback to the measurement-based feedback approach (see Figure~\ref{fig:Cooling_CFB_Compare_MFB}). For the measurement based feedback cooling, we adapt the results from~\cite{Genes2008} with the regime $\kappa_\OMC \gg \omega_\mathrm{fb} \sim \Omega_\Mech \gg \Gamma_\Mech$, where $\omega_\mathrm{fb}$ is the feedback bandwidth. We minimize the phonon number with respect to $\omega_\mathrm{fb}$ and the feedback gain numerically. We use a detection efficiency that matches the single-way optical path efficiency, $\eta_\mathrm{det} = \eta_\STrip$. At small quantum cooperativity, the measurement-based feedback cooling is much more efficient, as the weak cooling power for the coherent feedback cooling is due to the lack of gain in the coherent feedback system proposed here. On the other hand, in the measurement-based feedback cooling, the gain can be tuned to reach the noise-squashing regime~\cite{Schaefermeier2016,Rossi2018}. However, at large intra-cavity photon number, it is possible to achieve a lower phonon occupation with the coherent feedback. For the coherent feedback, the feedback control signal $\hat u_\mathrm{fb}$ coherently mixes with the vacuum noise ($\hat u_\mathrm{vac}$) with an efficiency $\eta$ coupling into the optomechanical cavity, $\hat u_\OMC^{\inp} = \sqrt{\eta} \hat u_\mathrm{fb} + \sqrt{1 - \eta} \hat u_\mathrm{vac}$. For the measurement-based feedback cooling, the control signal is classical, $\hat u_\OMC^{\inp} = \sqrt{\eta} u_\mathrm{fb} + \hat u_\mathrm{vac}$. The input noise is therefore lower in the coherent feedback case, with $0 < 1-\eta < 1$ always satisfied.

\subsection{Entanglement generation and verification between photons and phonons}
\label{ss:Results:Entanglement}

\begin{figure}[!t]
    \centering
    \includegraphics[width=1.\columnwidth]{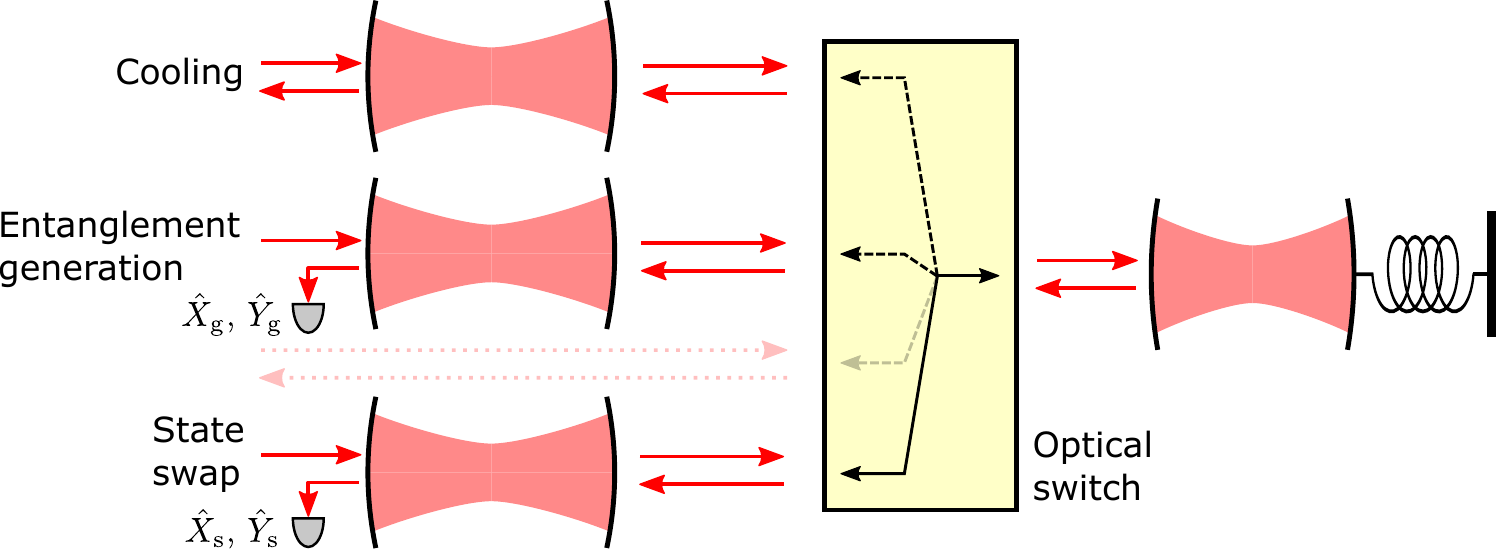}
    \caption{Entanglement generation and verification scheme considered in this work. The optomechanical cavity connects to three auxiliary cavities via an optical switch. The three cavities are for cooling, entanglement generation, and state-swap for entanglement verification. By measuring the output of the last two auxiliary cavities, photon-phonon entanglement can be detected. Between the entanglement generation and verification, a gap without the feedback is inserted. Optionally, as considered in this work, light from the laser is coupled into the optomechanical directly to keep a constant cavity photon number.}
    \label{fig:EntanglementScheme}
\end{figure}

Quantum entanglement between photons and phonons has been proposed using an optomechanical system in the sideband-resolved regime~\cite{Hofer2011,Galland2014} and experimentally demonstrated in various systems~\cite{Palomaki2013,Riedinger2016}. Depending on the detuning of the input field with respect to the optomechanical cavity, either the Stokes process can be used to generate entanglement or the anti-Stokes process serves as a readout and to verify the entanglement (state-swap operation). With a sideband-resolved system, the other process can always be strongly suppressed, which makes it possible to efficiently generate and verify entanglement by sending two pulses. In contrast, in a sideband-unresolved system, the difference in the suppression is lacking. Entanglement creation and the readout happens at a similar rate, making experimental implementations challenging. Inferring entanglement through continuous measurement inspired by the similar idea has been proposed, with a maximum squeezing of the EPR quadratures reaching up to 50\% of the vacuum noise~\cite{Gut2020}. The coherent feedback proposed can effectively bring the sideband suppression back and it is again possible to create and verify optomechanical entanglement with the help of an auxiliary cavity. Squeezing of the EPR quadratures beyond 50\% of the vacuum noise could be achieved by optimizing the mechanical quality factor with a reasonably high efficiency of the optical path.

We consider the scheme shown in Figure~\ref{fig:EntanglementScheme}, without any delay in the feedback to reduce the complexity of the model. It expands the coherent feedback scheme in Figure~\ref{fig:GeneralScheme} to a setup with three cavities. A fast optical switch allows to select which of the auxiliary cavities couples to the optomechanical system. Light is sent into the system through port 2 (the port connecting to the outside) of the auxiliary cavities. To start, the switch is set to the first cavity, which is used to pre-cool the mechanical resonator into a low thermal occupation state. For simplicity, we set $\kauxa/2\pi=400$~kHz, $\kauxb/2\pi = 100$~kHz, $\Delta_\aux=-\Omega_\Mech$, and $\phi_\STrip = 0$. We also use an interaction time of 0.1~s, which is sufficient to reach a steady state. We then switch to the second cavity, which is used for the entanglement generation. Finally, the third cavity is used to perform a state-swap in order to read the state of the mechanics. We assume an experimentally achievable switching time of 100~ns between the entanglement generation and the state-swap stage, during which there is no feedback (shown as an empty channel with dotted optical lines in Figure~\ref{fig:EntanglementScheme}). Our approach expands on the scheme presented in~\cite{Hofer2011}. By introducing the switching time between the second and the third cavity, which is much larger than $1/\kappa_\OMC$, the entanglement generation cavity and the state-swap cavity are effectively isolated. The light carrying information during the entanglement generation cannot be detected by the state-swap measurement. Optomechanical entanglement can thus be verified by measuring the $X$ and $Y$ quadratures of the output of the two cavities. 

We calculate the entanglement through the linearized dynamics described in Equation~\eqref{eq:CFBLangevinForm}, with $D=A_1=C_1 = 0$. The calculation routine is similar to that in reference~\cite{Rakhubovsky2015}, but without the use of the rotating wave approximation (RWA) for the mechanical resonator. The evolution of the system is of the form
\begin{equation}
    \hat u (t) = \exp(A_0 t) \hat u(t_0) + \int_{t_0}^t \dd s \exp \left(A_0 (t_0-s)\right) C_0 \hat u_\inp(s).
\end{equation}
For $t$ being in different stages (pre-cooling, entanglement generation, switching, state-swap), $t_0$ represents the start of each stage. With this setting, $A_0$ and $C_0$ are constant. The output of the auxiliary cavities, for the entanglement generation and verification, are then defined as
\begin{equation}
    \begin{gathered}
        \hat u_\alpha = u_{\aux, \alpha}^{\inp,2} - \sqrt{\kappa_{\aux,\alpha}} u_{\aux,\alpha}.
    \end{gathered}
\end{equation}
$\alpha \in \{\mathrm{g, s}\}$ denotes the components involved in the entanglement generation and the state-swap phase. Further, we define optical temporal modes~\cite{Hofer2011,Gut2020}
\begin{equation}
    \hat r_\alpha = \int_{t_0}^{t_f} \dd t f_\alpha(t) R(\theta_\alpha(t)) \hat u_\alpha.
\end{equation}
The integration is carried out only within the corresponding stage, starting from $t_0$ and ending at $t_f$. $R$ is the rotation matrix defined in Equation~\eqref{eq:RotMat}, with $\theta_\tmwrite (t) = \Omega_\Mech t$ and $\theta_\tmread (t) = - \Omega_\Mech t + \phi_\tmread$. The rotation matrix is necessary since we do not use the RWA for the sideband-unresolved regime. Also, we take the exponential form for the envelope $f_\alpha$~\cite{Hofer2011,Gut2020}
\begin{equation}
    \begin{gathered}
        f_\tmwrite (t) = \left(\frac{1 - \ee^{-2 \Gamma_\tmode \tau_\pulse}}{2 \Gamma_\tmode}\right)^{1/2} \ee^{\Gamma_\tmode (t - t_\tmwrite^{\mathrm{(f)}})}, \\
        f_\tmread (t) = \left(\frac{1 - \ee^{-2 \Gamma_\tmode \tau_\pulse}}{2 \Gamma_\tmode}\right)^{1/2} \ee^{- \Gamma_\tmode (t - t_\tmread^{\mathrm{(0)}})}.
    \end{gathered}
\end{equation}
$t_\tmwrite^{\mathrm{(f)}}$ is the end time of the entanglement generation, and $t_\tmread^{\mathrm{(0)}}$ is the starting time of the state-swap process. They can be chosen to be centered around $t=0$ ($t_\tmwrite^{\mathrm{(f)}} = - t_\tmread^{\mathrm{(0)}}$). $\Gamma_\tmode$ is a parameter controlling the exponential decay rate of the envelop. Both processes have a duration of $\tau_\pulse$. Outside the period they are equal to 0. This definition ensures that the temporal mode $\hat r$ satisfies the bosonic commutation relation, $[\hat r_{\alpha,i},~ \hat r_{\beta,j}] = \delta_{\alpha \beta} \epsilon_{ij}$, where $i, j \in \{X, Y\}$ are for the two quadratures included in $\hat r$. The covariance matrix of the temporal mode can then be evaluated
\begin{equation}
    \sigma_{ij} = \langle \hat r_i \hat r_j + \hat r_j \hat r_i \rangle,
\end{equation}
with $\hat r = (\hat r_{\tmwrite, X}, ~ \hat r_{\tmwrite, Y}, ~ \hat r_{\tmread, X}, ~ \hat r_{\tmread, Y})$. In this work, we use the EPR-variance to quantify the entanglement, $\Delta_\mathrm{EPR} = (\sigma_{11} + \sigma_{22} + \sigma_{33} + \sigma_{44})/2 + (\sigma_{13} - \sigma_{24})$~\cite{Gut2020}.

We note that switching with a finite dead time might introduce classical noise to the mechanical resonator due to the resulting change in the cavity photon number. It thus reduces the entanglement and is not captured by the linearized model~\cite{Teh2017}. It is more relevant to the calculation here, as the mechanical oscillation considered is significantly less coherent compared to the typical optomechanical experiments in the sideband-resolved regime~\cite{Palomaki2013,Wallucks2020}. A higher interaction strength with a shorter pulse is favorable, as demonstrated below. However, we stress that the switching time is much shorter than the mechanical oscillation period and thus the disturbance would mainly be at very high frequency, while the impact on a low-frequency mechanical resonator is minimal. Assuming an unchanged average photon number inside the optomechanical cavity, allows to eliminate the effect of the switching. Experimentally this can be realized by introducing another coupling channel, such as another waveguide into the optomechanical system. Alternatively, an additional channel on the optical switch that couples to the laser directly can be introduced. We include this in the scheme shown in Figure~\ref{fig:EntanglementScheme}, plotted as the dotted red lines for the optics and the dashed line inside the switch. By controlling the light intensity on this additional channel, it is then possible to achieve a constant cavity photon number. This approach reduces the entanglement since the photons during the switching are not measured, corresponding to a loss of information. Meanwhile, the photons interact with the mechanical resonator, creating mechanical decoherence. Therefore, a switch with a short switching time is required to reduce the switching impact. We note that on-chip optical switches with low loss and short switching time have been demonstrated experimentally~\cite{Abdalla2004,Sun2018}.

\begin{figure}[!t]
    \centering
    \includegraphics[width=1.\columnwidth]{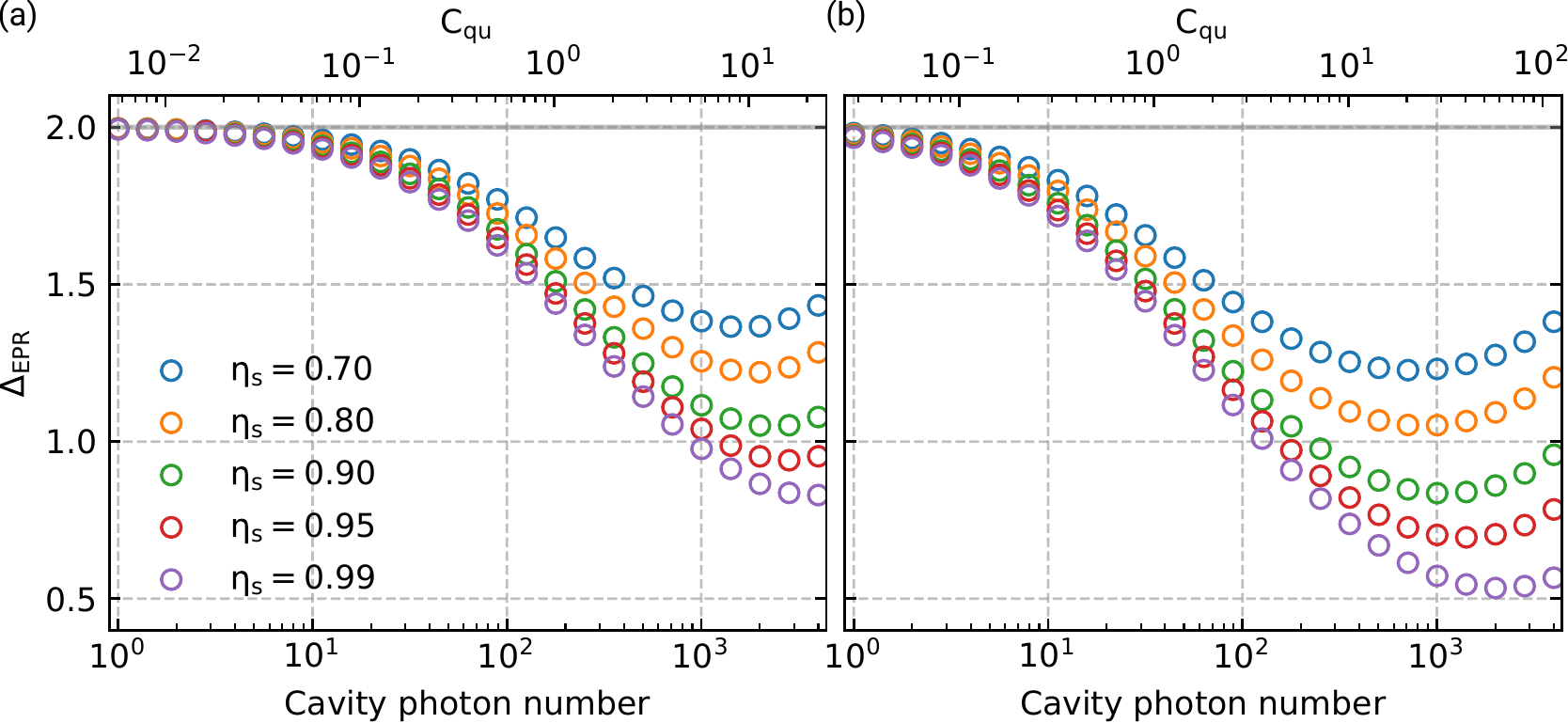}
    \caption{Optimized $\Delta_\mathrm{EPR}$ for (a) $Q_\Mech = 2\times 10^7$ and (b) $Q_\Mech = 10^8$. The gray solid line shows the separability bound $\Delta_\mathrm{EPR} = 2$.}
    \label{fig:EntanglementDeltaEPR}
\end{figure}

In the scheme considered here, the auxiliary cavities for entanglement generation and verification have the same parameters, except for an opposite detuning. The phases of the optical paths $\phi_\STrip$ for both stages also have opposite signs. We then minimize $\Delta_\mathrm{EPR}$ with respect to $\kauxa$, $\kauxb$, $\Delta_\aux$, $\phi_\STrip$, $\Gamma_\tmode$ and $\tau_\pulse$, where the values are given in the appendix (Figure \ref{fig:DeltaEPROptimParam}). Different from the feedback cooling, we include $\kauxb$ as an optimization parameter since the entanglement is detected from the output light of channel 2. The result is plotted in Figure~\ref{fig:EntanglementDeltaEPR}, with a unity detection efficiency for the light getting out of the feedback system. We consider mechanical resonators with $Q_\Mech = 2\times 10^7$ and $10^8$, corresponding to a thermal decoherence rate, in the unit of mechanical resonance cycles, $Q_\Mech f_\Mech / (k_\mathrm{B} T / h)$ of 230 and 1100, respectively. When increasing the cavity photon number, we can observe a reduction in $\Delta_\mathrm{EPR}$. For a single-way efficiency of 0.7, 0.8 and 0.9, it is possible to achieve a $\Delta_\mathrm{EPR}$ of 1.37, 1.22 and 1.05 with $Q_\Mech = 2\times 10^7$, and 1.23, 1.05 and 0.84 with $Q_\Mech = 10^8$, respectively. $\Delta_\mathrm{EPR} < 2$ is the threshold for entanglement between the temporal modes of the two output light pulses and is a direct result of the entanglement between the photons and the motion of the mechanical resonator. Increasing the photon number beyond an optimal point leads to an increase in $\Delta_\mathrm{EPR}$. In the present of optical loss, a stronger interaction leads to a stronger effective noise, reducing $\tilde C_\mathrm{qu}$. As the interaction strength is increased, the time of the pulse becomes shorter. The fixed switching time, during which the entanglement is reduced, becomes therefore more dominant. The calculation shows that it is beneficial to achieve low loss over the feedback path and a low thermal decoherence. Still, it is remarkably robust against any dissipation in the system, making it feasible for real experimental parameters. In the optimization, an optimal $\kauxb/2\pi$ is around 500~kHz. We note that a total linewidth of 220~kHz has been reported recently using on-chip disk resonators~\cite{Wu2020b}, which is promising for a fully integrated coherent feedback system, which could minimize optical loss over the feedback path. A reasonable increase of the mechanical frequency would also make a fully integrated system more feasible.\\

\section{Conclusion}

In this work, we propose a coherent feedback scheme with linear, passive optical components. We mainly consider optomechanical systems in the deep sideband-unresolved regime, and with experimentally relevant parameters. We show that an additional, external optical cavity can effectively bring the optomechanical system into the sideband-resolved regime for a specific set of parameters ($\phi_\STrip=0$, $\tau_\STrip=0$, $\Delta_\OMC = 0$). We consider non-unity feedback efficiency, which introduces additional noise to the mechanical resonator. Overall, the effective quantum cooperativity can still be enhanced, depending on the feedback path efficiency and the original quantum cooperativity. Our analysis shows that coherent feedback is a highly promising path for broad applications using sideband-unresolved systems.

We use these results to demonstrate how either an optical cavity or a mirror plus an optical delay path as an auxiliary component can be used to perform groundstate cooling of the mechanical resonator under practical experimental conditions. Furthermore, based on an entanglement protocol with long pulses~\cite{Hofer2011}, we then propose an experimental scheme that uses three auxiliary cavities for cooling, entanglement generation and verification. By switching between these cavities in a relatively short time the output light can be used to detect photon-phonon entanglement. We quantify the entanglement of the output light by evaluating the EPR-variance of the temporal optical mode. Even though it is not necessarily the optimal entanglement witness~\cite{Hyllus2006,Gut2020} it shows a significant squeezing well below the inseparability bound. Experimentally realizing a fully integrated on-chip coherent feedback structure is within reach of state-of-the-art on-chip optical resonators~\cite{Wu2020b,Puckett2021} and fast optical switches~\cite{Abdalla2004,Sun2018}, even for mechanical frequency as low as 1~MHz. Such an integrated structure would drastically reduce the complexity of an experiment and could help realize novel quantum applications~\cite{Barzanjeh2022} with optomechanical systems in the sideband-unresolved regime.

\begin{acknowledgments}
    We would like to thank Jie Li and Corentin Gut for valuable discussions. This work is supported by the European Research Council (ERC CoG Q-ECHOS, 101001005), and by the Netherlands Organization for Scientific Research (NWO/OCW), as part of the Frontiers of Nanoscience program, as well as through a Vrij Programma (680-92-18-04) grant. J.G.\ gratefully acknowledges support through a Casimir PhD fellowship.
\end{acknowledgments}

\section*{Data Availability}
Source data for the plots are available on \href{https://doi.org/10.5281/zenodo.7243578}{Zenodo}.

\clearpage

\appendix

\section{Default parameters and convention}\label{sect:DefaultParams}
If not specified otherwise, we consider the following parameters.

\vspace{0.5cm}

\begin{widetext}
\begin{center}
\begin{tabular}{ || c | c || }
    \hline
    Mechanical frequency $\Omega_\Mech/2\pi$ & 1 MHz \\ \hline
    Mechanical quality factor $Q_\Mech=\Omega_\Mech/\Gamma_\Mech$ & $2\times 10^7$ \\ \hline
    Energy decay rate of optomechanical cavity $\kappa_\OMC$ & $2\pi\times10~$GHz \\ \hline
    Detuning of the optomechanical cavity $\Delta_\OMC$ & 0 \\ \hline
    Coupling efficiency of optomechanical cavity $\eta_\OMC = {\kOMe} / {\kappa_\OMC}$ & 0.8 \\ \hline
    Coupling rate of auxiliary cavity to the feedback $\kauxa$ & $2\pi\times400~$kHz \\ \hline
    Coupling rate of auxiliary cavity to other channels $\kauxb$ &  $2\pi\times100~$kHz \\ \hline
    Environment temperature $T$ & 4.2 K \\ \hline
    Delay (only for the feedback by auxiliary cavity) $\tau_\STrip$ & 0 \\ \hline
    ``Switching time'' of the temporal mode function $t_\tmread^{\mathrm{(0)}}-t_\tmwrite^{\mathrm{(f)}}$ & 0.1~$\upmu$s \\ \hline
\end{tabular}
\end{center}\end{widetext}

We define the Fourier transform with the convention
\begin{equation}
	u(\omega) = \mathcal{F}[u (t)] (\omega) = \int_{-\infty}^{+\infty} u(t) \ee^{\I\omega t} \dd t.
\end{equation}

\section{Steady state and photon number}\label{sect:SteadyState}
When performing the feedback cooling, the steady state of the system is considered. It can be analyzed by transforming Equation~\eqref{eq:CFBLangevinForm} to the Fourier domain,
\begin{equation}\label{eq:CFBLangevinFormFourier}
    \begin{aligned}
        &-\I \omega \left( I + D  \ee^{\I \omega \tau}\right) \hat u (\omega) = \\
        & \left( A_0 + A_1 \ee^{\I\omega\tau}\right) \hat u(\omega) + \left(\sum_{n=0}^{2} C_n  \ee^{\I n\omega \tau_\STrip} \right) \hat u_\inp(\omega),
    \end{aligned}
\end{equation}
where $I$ is the identity matrix. Rearranging Equation~\eqref{eq:CFBLangevinFormFourier} yields the form $\hat u(\omega) = M(\omega) \hat u_\inp(\omega)$, where $M(\omega)$ is the transfer matrix
\begin{equation}
    \begin{aligned}
        M(\omega) = - \left( \I \omega \left( I + D \ee^{\I\omega \tau} \right) + A_0 + A_1 \ee^{\I\omega\tau} \right)^{-1} \\ \times \left(\sum_{n=0}^{2} C_n  \ee^{\I n\omega \tau_\STrip} \right).
    \end{aligned}
\end{equation}
The input noise has a single-side spectrum~\cite{Genes2008}
\begin{equation}
    \begin{gathered}
        S_{{u_\inp}_i}(\omega) = 1, \\
        S_{X_\Mech^\inp}(\omega) = S_{Y_\Mech^\inp}(\omega) = 2 n_\therm + 1.
    \end{gathered}
\end{equation}
Here, ${{u_\inp}_i}$ are for the elements corresponding to the optical input noise only. It is then possible to get the spectrum of $\hat u$~\cite{Genes2008},
\begin{equation}\label{eq:theory_tf_u_uinp}
    S_{u} (\omega) = \left| M (\omega) \right|^2 S_{u_\inp}.
\end{equation}
The absolute value and the square are performed entry-wise. This allows extracting the energy of the mechanical resonator by integrating the spectrum of the mechanical field. The corresponding phonon occupancy is given by~\cite{Genes2008},
\begin{equation}\label{eq:CFB_theory_nphn}
    n_\mathrm{phn} = \frac{1}{2} \left(\int_0^{\infty} \frac{d \omega}{2\pi} \left( S_{X_\Mech} (\omega) + S_{Y_\Mech} (\omega) \right)\right) - \frac{1}{2}.
\end{equation}

The scheme is valid only when the system is stable. Determining the stability of the system can be done in a classical way~\cite{Genes2008}. For our system with delay, we follow the method described in ~\cite{Louisell2001,Olgac2004} to perform the stability test.

\clearpage

\section{Expression of the effective fields in Equation~(\eqref{Eq:sim_single_cav_Langevin})}
\label{app:eff_field_sim_single_cav}

\begin{widetext}\begin{equation}
        \begin{gathered}
            \hat {\tilde u}_\aux ^{\inp,1} = \frac{1}{\sqrt{\xi_2}} \left(- \frac{\sqrt{\eta_\STrip \kOMe \kOMi}}{\kappa_\OMC/2} \hat u_\OMC^{\inpi} + r_\OM \sqrt{\eta_\STrip (1-\eta_\STrip) \hat u_\backw^{\inp}} + \sqrt{1-\eta_\STrip} \hat u_\forw^\inp \right), \\
            \hat u_\add = \sqrt{\frac{\eta_\OM}{(1-\eta_\STrip) \xi_1}} \left( (1-\eta_\STrip) \sqrt{\frac{\kOMi}{\kOMe}} \hat u_\OMC^{\inpi} + \sqrt{1-\eta_\STrip} u_\backw^{\inp} + \sqrt{\eta_\STrip (1-\eta_\STrip)} u_\forw^{\inp} \right).
        \end{gathered}
\end{equation}\end{widetext}
They are a result of the combination of the optical vacuum field, and they satisfies the Bosonic commutation relation.

\vspace{0.5cm}

\section{Optimized parameters for Figure~\ref{fig:Cooling_CR_tau_0} and Figure~\ref{fig:EntanglementDeltaEPR}}

\begin{figure}[H]
    \centering
    \includegraphics[width=\columnwidth]{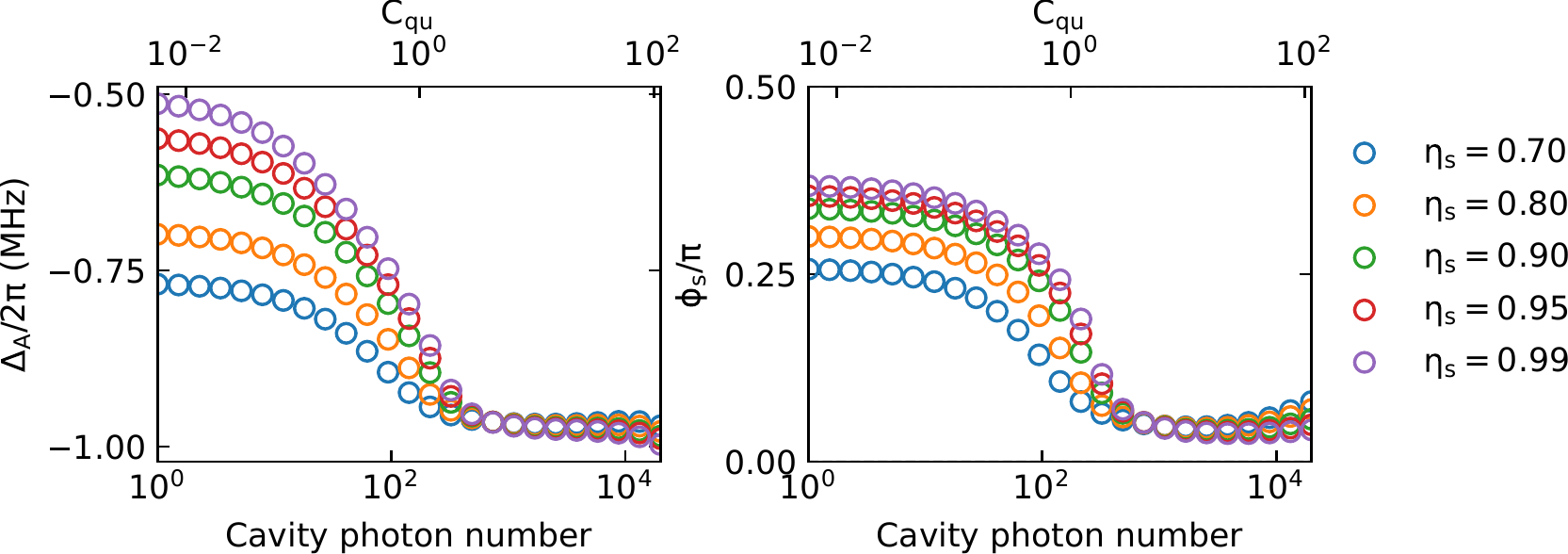}
    \caption{Optimized parameters used in Figure~\ref{fig:Cooling_CR_tau_0}(a).}
    \label{fig:Cooling_CR_tau_0_supp}
\end{figure}

\begin{figure*}[h!]
    \centering
    \includegraphics[width=\textwidth]{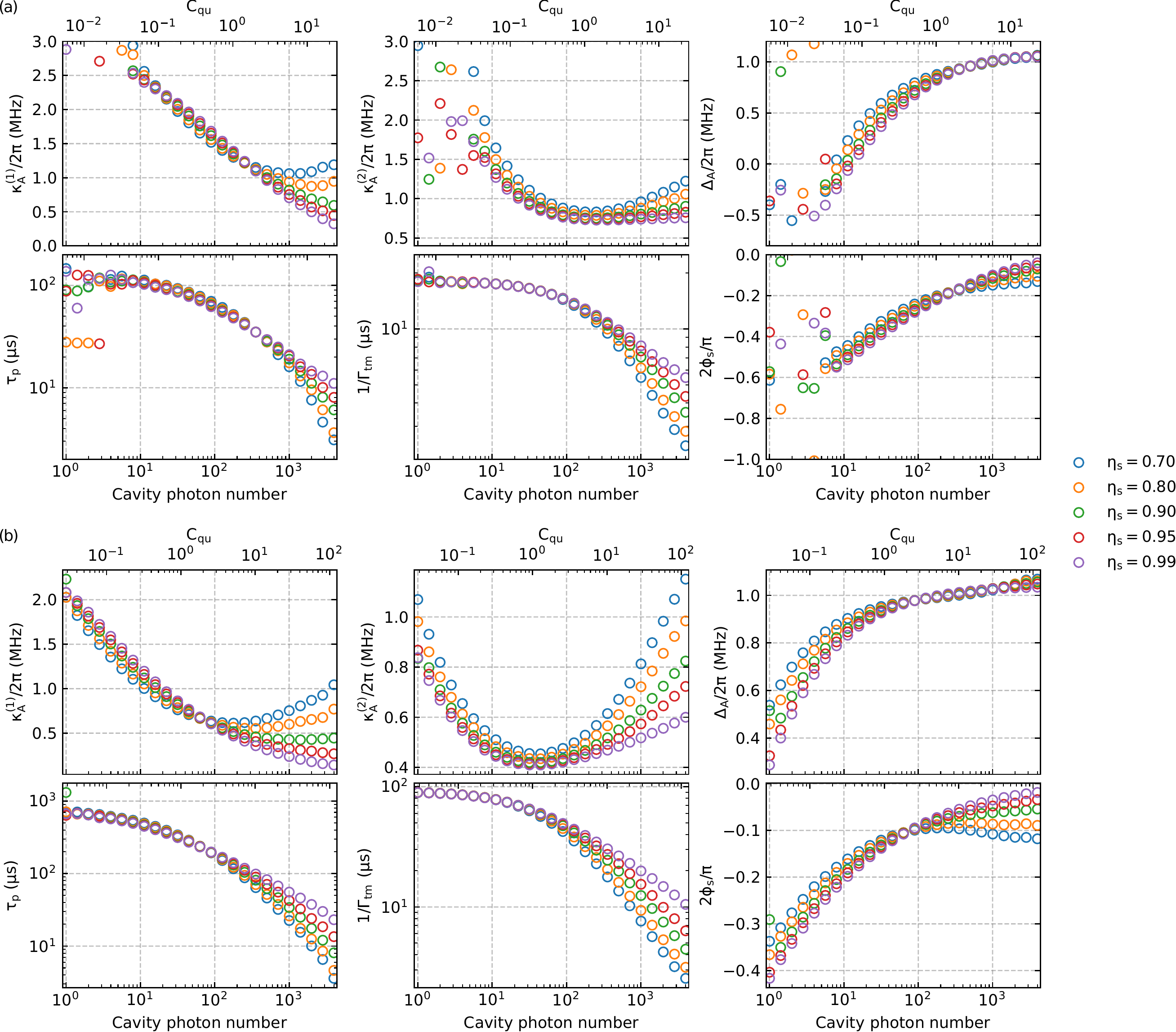}
    \caption{Optimized parameters in Figure~\ref{fig:EntanglementDeltaEPR} for (a) $Q_\Mech = 2 \times 10^7$ and (b) $Q_\Mech = 10^8$.}
    \label{fig:DeltaEPROptimParam}
\end{figure*}

\FloatBarrier

\end{document}